\documentstyle[preprint,aps,epsfig]{revtex}
 
\begin{document}
 \draft
\title{Rapid cooling of magnetized neutron stars}
\author{Debades Bandyopadhyay,$^1$ Somenath Chakrabarty,$^{2,3}$ 
Prantick Dey,$^2$ and Subrata Pal$^1$ }
\address{$^1$Saha Institute of Nuclear Physics, 1/AF Bidhannagar, 
Calcutta 700 064, India}
\address{$^2$Department of Physics, University of Kalyani, 
Kalyani 741235, India} 
\address{$^3$IUCAA, P.B. 4, Ganeshkhind, Pune 411007, India}

\maketitle

\begin{abstract}
The neutrino emissivities resulting from direct URCA processes in 
neutron stars are calculated in a relativistic Dirac-Hartree approach 
in presence of a magnetic field. In a quark or a hyperon matter environment, 
the emissivity due to nucleon direct URCA processes is suppressed relative 
to that from pure nuclear matter. In all the cases studied, the magnetic 
field enhances emissivity compared to the field-free cases. 
\end{abstract}


Neutron stars are born in the aftermath of supernova explosions with 
interior temperatures $T\stackrel{>}{\sim} 10^{11}$ K, but cool 
rapidly in a few seconds by predominant neutrino emission \cite{Bur} 
to $T<10^{10}$ K.  Neutrino cooling then dominates and lasts for 
$t\sim 10^5-10^6$ yr and subsequently photon emission takes over when 
$T\stackrel{<}{\sim} 10^8$ K. Since the long term cooling of the young 
neutron stars ($T\sim10^8-10^{10}$ K) proceeds via emission of neutrinos 
primarily from matter at supranuclear densities within the core, the study 
of the cooling of neutron stars by examination of neutrino emissivities may 
provide considerable insight into their interior structure and composition.

For a long time the dominant neutrino cooling mechanism has been the 
so-called standard model based on the modified URCA processes \cite{Chi,Bah}
\begin{eqnarray}
(n,p) + n &\to& (n,p) + p + e^- + {\overline\nu}_e, \nonumber\\
(n,p) + p + e^- &\to& (n,p) + n + \nu_e.
\end{eqnarray}
The ROSAT detection \cite{Oge} of thermal emission from neutron stars
indicates the necessity of faster cooling mechanism in some young neutron
stars, in particular the Vela pulsar. Faster neutrino emission than the 
standard model was proposed by invoking pion \cite{Bah,Max} or kaon \cite{Bro} 
condensates which have neutrino emissivities comparable to that from 
the $\beta$-decay of quarks \cite{Iwa} in quark matter (consisting of 
$u$, $d$ and $s$ quarks)
\begin{equation}
d \to u + e^- + {\overline\nu}_e, \;\;\;  u + e^- \to d + \nu_e.
\end{equation}
A similar relation for $s$-quark $\beta$-decay may occur and is obtained 
by replacing $d$ by $s$ quark in Eq. (2). The most powerful energy losses, 
expected to date, are produced by the so-called direct URCA mechanism 
involving nucleons \cite{Lat}
\begin{equation}
n \to p + e^- + {\overline\nu}_e, \;\;\;  p + e^- \to n + \nu_e.
\end{equation}
The threshold density of this process is however considerably larger than 
that of the modified URCA process.

Observations of pulsars predict large surface magnetic field of 
$B_m \sim 10^{14}$ G \cite{Chanm}. In the core the field may be considerably
amplified due to flux conservation from the original weak field of the 
progenitor during its core collapse. In fact, the scalar virial theorem
\cite{Chand} predicts large interior field $B_m \sim 10^{18}$ G or more 
\cite{Chak}, and these fields are frozen in the highly conducting core. 
It has been demonstrated \cite{Chak} that when the field $B_m$ is comparable 
to or above a critical field $B_m^{(c)}$, the energy of a charged particle 
changes significantly in the quantum limit; the quantum effects are most 
pronounced when the particle moves in the lowest Landau level. The phase 
space modifications stemming from the strong magnetic field in the core are 
expected to influence the neutrino emission rate from young neutron stars.

In this communication we evaluate the neutrino emissivity for the nucleon 
direct URCA process of Eq. (3) in presence of a magnetic field $B_m$, and
demonstrate that it would lead to more rapid cooling in the core. 
(A straightforward extension of the emissivity for nucleons into the
quark sector may also be obtained in a magnetic field.) For this purpose,
we consider a $npe$ matter in $\beta$-equilibrium within a relativistic
Dirac-Hartree approach in the linear $\sigma$-$\omega$-$\rho$ model \cite{Ser}.

At the neutron star core at temperatures well below the typical
Fermi temperature of $T_F\sim 10^{12}$ K, the nucleons and electrons
participating in neutrino producing processes are all degenerate 
(the $\nu_e$ and ${\overline\nu}_e$ are free) and have their momenta
close to the Fermi momenta $p_{F_i}$, where $i=n,p,e$. Since neutrino 
and antineutrino momenta are $\sim kT/c \ll p_{F_i}$, the nucleon direct
URCA process is allowed by the momentum conservation when 
$p_{F_p} + p_{F_e} \geq p_{F_n}$. Since matter is very close to 
$\beta$-equilibrium, the chemical potentials of the constituents satisfy the
condition $\mu_n=\mu_p+\mu_e$. (Henceforth we set $\hbar=c=k=1$.) 

Employing the Weinberg-Salam theory for weak interactions, the interaction
Lagrangian density for the charged current reaction (3) may be expressed as
${\cal L}_{\rm int}^{\rm cc} = (G_F/{\sqrt 2}) \cos\theta_c l_\mu j^\mu_W$, 
where $G_F \simeq 1.435\times 10^{-49}$ erg cm$^{-3}$ is the Fermi
weak coupling constant and $\theta_c$ the Cabibbo angle.
The lepton and nucleon charged weak currents are respectively, 
$l_\mu = {\overline \psi}_4 \gamma_\mu (1-\gamma_5)\psi_2$ and
$j_W^\mu = {\overline \psi}_3 \gamma^\mu (g_V-g_A\gamma_5)\psi_1$. Here,
and in other formulae to follow, the indices $i=1-4$ refer to the $n$,
${\overline \nu}_e$, $p$ and $e$, respectively. The vector and axial-vector 
coupling constants are $g_V=1$ and $g_A=1.226$. 

The emissivity due to the antineutrino emission process in presence
of a uniform magnetic field $B_m$ along z-axis when both the electrons 
and protons are Landau quantized is given by
\begin{eqnarray}
\varepsilon_\nu(B_m) &=& 2 \int \frac{Vd^3p_1}{(2\pi)^3} 
\int \frac{Vd^3p_2}{(2\pi)^3}
\int^{qB_mL_x/2}_{-qB_mL_x/2} \frac{L_ydp_{3y}}{2\pi}
\int^{p_{F_p}}_{-p_{F_p}} \frac{L_zdp_{3z}}{2\pi}
\int^{qB_mL_x/2}_{-qB_mL_x/2} \frac{L_ydp_{4y}}{2\pi}
\int^{p_{F_e}}_{-p_{F_e}} \frac{L_zdp_{4z}}{2\pi} \nonumber \\
&\times& \sum_{\eta=0}^{\eta_{\rm max}}
\sum_{\eta^{'}=0}^{\eta^{'}_{\rm max}}
E_2 W_{fi} f({\bf p}_1) [1-f({\bf p}_3)] [1-f({\bf p}_4)] ~,
\end{eqnarray}
where $\eta_{\rm max}$ and $\eta^{'}_{\rm max}$ are respectively the maximum 
number of Landau levels populated for protons and electrons. The prefactor
2 takes into account the neutron spin degeneracy. The 
$p_i\equiv (E_i,{\bf p}_i)$ are the 4-momenta and $E_2$ the antineutrino 
energy. The functions $f(E_i)$ denote the Fermi-Dirac functions for the 
$i$th particle. The transition rate per unit volume due to the antineutrino 
emission process may be derived from Fermi's golden rule and is 
given by $W_{fi}=\langle|{\cal M}_{fi}|^2\rangle/(tV)$.
Here $t$ represents time and $V=L_xL_yL_z$ the normalization volume. 
$|{\cal M}_{fi}|^2$ is the squared matrix element and the symbol 
$\langle \cdot\rangle$ denotes an averaging over initial spins and a sum 
over final spins. The matrix element for the $V-A$ interaction is given by
\begin{equation}
{\cal M}_{fi} = \frac{G_F}{\sqrt 2} \int d^4X \: {\overline \psi}_1(X)
\gamma^\mu \left(g_V-g_A\gamma_5\right) \psi_3(X) \:
{\overline \psi}_2(X)\gamma_\mu \left(1-\gamma_5\right)\psi_4(X) ~.
\end{equation}
In presence of a uniform magnetic field $B_m$, the normalized proton 
wave function is $\psi_3(X) = \left(1/\sqrt{L_yL_z} \right) 
\exp\left(-i E_3 t + ip_{3y}y + ip_{3z} z\right) f_{p_{3y},p_{3z}}(x)$,  
where $f_{p_{3y},p_{3z}}(x)$ is the 4-component spinor solution \cite{Chak}.
The form of the spinor in a magnetic field (see Ref. \cite{Chak}) restricts
the analytical evaluation of the neutrino emissivity to fields strong enough
so as to populate only the ground state for electrons and protons,
i.e. $\eta=\eta^{'}=0$. The only positive energy spinor for protons
in the chiral representation is then \cite{Chak,Ban}
\begin{equation}
f^{\eta=0}_{p_{3y},p_{3z}}(x) = N_{\eta=0} \left( \matrix{ 
E_3^* + p_{3z}\cr  
0\cr  - m^*\cr  0\cr} \right) I_{{\eta=0};p_{3y}}(x) , 
\end{equation}
where $N_{\eta=0} = 1/\sqrt{2E_3^*(E_3^* + p_{3z})}$, 
and $E_3^* = E_3 - U^H_{0;p} = (p_{3z}^2+{m^*}^2)^{1/2}$ 
is the effective relativistic Hartree energy. The function 
$I_{\eta=0;p_{3y}}(x)$ is similar in form as in Ref. \cite{Chak}.
The nucleon effective and rest masses are respectively, $m^*$ and  
$m = m_n = m_p = 939$ MeV. In presence of the magnetic field,
the wave functions for free electrons $\psi_4(X)$ have the same 
form as those for protons, but with $m^*$ and $E_3$ 
for protons replaced by the bare mass $m_e$ and kinetic energy for electrons,
respectively. The neutrons and neutrinos/antineutrinos being unaffected 
by $B_m$, have plane wave functions.

Using these wave functions, it is straightforward to calculate the 
transition rate per unit volume and is given by
\begin{eqnarray}
W_{fi} &=& \frac{G_F^2}{E_1^*E_2E_3^*E_4} \: \frac{1}{V^3L_yL_z}
\exp\left[- \frac{(p_{1x}-p_{2x})^2 + (p_{3y}+p_{4y})^2}{2qB_m}\right] 
\nonumber\\
&\times& \left[(g_V + g_A)^2(p_1\cdot p_2)(p_3\cdot p_4)
+ (g_V - g_A)^2(p_1\cdot p_4)(p_3\cdot p_2)
- (g_V^2 - g_A^2)m^{* 2}(p_4\cdot p_2) \right] \nonumber\\
&\times&\delta(E_1-E_2-E_3-E_4) 
\delta(p_{1y}-p_{2y}-p_{3y}-p_{4y}) 
\delta(p_{1z}-p_{2z}-p_{3z}-p_{4z}) .
\end{eqnarray}
Substituting Eq. (7) in the expression (4) for emissivity, and by the 
change of variable $(p_{3y}+p_{4y}) \to p_{3y}$, the integration over
$dp_{4y}$ can be performed to yield a factor $qB_mL_x$. The rest of 
the integrals of Eq. (4) can then be performed in the standard manner 
\cite{Iwa}. Electron capture gives the same emissivity as neutron decay, 
although in neutrinos, and thus the total emissivity (relativistically) 
for the direct URCA process in nuclear matter (NM) in a magnetic field 
$B_m$ is $\varepsilon^{\rm NM}_{\rm URCA}(B_m) = 2 \varepsilon_\nu(B_m)$ i.e.
\begin{eqnarray}
\varepsilon^{\rm NM}_{\rm URCA}(B_m) &=& \frac{457\pi}{5040} G_F^2 
\cos^2\theta_c \: (qB_m) 
\bigg[ \left(g_V + g_A\right)^2
\left(1-\frac{p_{F_p}}{\mu^*_p} \right) + \left(g_V-g_A\right)^2
\left(1-\frac{p_{F_n}}{\mu^*_n}\cos\theta_{\rm 14}\right) \nonumber\\
&-& (g_V^2-g_A^2) \frac{m^{* 2}}{\mu_n^*\mu_p^*} \Bigg] 
\exp\left[ \frac{(p_{F_p}+p_{F_e})^2-p_{F_n}^2}{2qB_m} \right] 
\frac{\mu^*_n \mu^*_p \mu_e} {p_{F_p} p_{F_e}} T^6 \Theta ,
\end{eqnarray}
where $\mu^*_i = (p_i^2+m^{* 2})^{1/2}$ and 
$\cos\theta_{\rm 14}=(p_{F_n}^2 + p_{F_e}^2 - p_{F_p}^2)/2 p_{F_n} p_{F_e}$. 
The threshold factor is
$\Theta = \theta(p_{F_p}+p_{F_e}-p_{F_n})$, where
$\theta(x)=1$ for $x>0$ and zero otherwise.
For $B_m=0$, the relativistic expression for the
neutrino emissivity from the nucleon direct URCA process is
\begin{eqnarray}
\varepsilon^{\rm NM}_{\rm URCA}(B_m=0) &=& \frac{457\pi}{10080} G_F^2 
\cos^2\theta_c \bigg[ \left(g_V + g_A\right)^2
\left(1-\frac{p_{F_p}}{\mu^*_p} \cos\theta_{\rm 34} \right) \nonumber\\
&+& \left(g_V-g_A\right)^2
\left(1-\frac{p_{F_n}}{\mu^*_n}\cos\theta_{\rm 14}\right)
-(g_V^2-g_A^2) \frac{m^{* 2}}{\mu_n^*\mu_p^*} \Bigg] 
\mu^*_n \mu^*_p \mu_e T^6 \Theta .
\end{eqnarray}

It was shown \cite{Iwa} that quark matter (QM), if present, the $\beta$-decay 
(i.e. direct URCA process) of $d$ quarks is kinematically allowed through 
reaction (2) if finite mass (and/or quark-quark interaction) is incorporated. 
The relativistic expression of the neutrino emissivity for the direct URCA 
process involving $u$ and $d$ quarks for $B_m=0$ and without 
quark-quark interaction is given by \cite{Iwa}
\begin{equation}
\varepsilon^{\rm QM}_{\rm URCA}(B_m=0) = \frac{457\pi}{840} G_F^2 
\cos^2\theta_c (1-\cos\theta_{\rm 34}) \mu_d \mu_u \mu_e T^6 ,
\end{equation}
in the usual notation \cite{Iwa}. The emissivity for the $\beta$ decay of 
$s$ quark for $B_m=0$ is similar to Eq. (10) with $\cos\theta_c$ replaced 
by $\sin\theta_c$. The emissivity for the $\beta$ decay of free $d$ quark 
in a magnetic field $B_m$ may be obtained from Eq. (8) by substituting 
$g_V=g_A=1$ with $\mu^*_n\to \mu_d$, $\mu^*_p\to \mu_u$, and multiplying a
color factor 3 for $d$ quark:
\begin{eqnarray}
\varepsilon^{\rm QM}_{\rm URCA}(B_m) &=& \frac{457\pi}{420} G_F^2 
\cos^2\theta_c \: (qB_m) \left(1-\frac{p_{F_u}}{\mu_u}\right) 
\exp\left[\frac{(p_{F_u}+p_{F_e})^2-p_{F_d}^2}{2qB_m} \right] 
\frac{\mu_d \mu_u \mu_e} {p_{F_u} p_{F_e}} T^6 .
\end{eqnarray}
Similar expression is obtained for $s$ quark, but is Cabibbo suppressed.
The decay of $d$ and $s$ quarks is feasible if they satisfy the respective 
inequality conditions $p_{F_u}-p_{F_e} \leq p_{F_d} \leq p_{F_u}+p_{F_e}$ 
and $p_{F_u} - p_{F_e} \leq p_{F_s} \leq p_{F_u} + p_{F_e}$. 

To estimate numerically the various neutrino emissivities for the direct
URCA processes with and without magnetic field in a neutron star, we describe
the nuclear matter and electrons within the relativistic Hartree approach
in the linear $\sigma$-$\omega$-$\rho$ model \cite{Chak,Ban}. 
The values for the dimensionless 
coupling constants for the $\sigma$, $\omega$ and $\rho$ mesons are 
adopted from Ref. \cite{Gle} which are determined by reproducing the nuclear
matter properties at a saturation density of $n_0=0.16$ fm$^{-3}$. The
variation of magnetic field with density $n_b$ from surface to center 
of the star is parametrized by the form \cite{Ban}
\begin{equation} 
B_m(n_b/n_0) = B_m^{\rm surf} + B_0 \left[
1 - \exp\left\{ -\beta (n_b/n_0)^{\gamma} \right\} \right] ,
\end{equation} 
where the parameters are chosen to be $\beta = 10^{-4}$ and $\gamma = 6$.
The maximum field prevailing at the center is taken as
$B_0=5\times 10^{18}$ G and the surface field is 
$B_m^{\rm surf} \simeq 10^{8}$ G. The number of Landau levels populated for 
a given species is determined by the $B_m$ and $n_b$ \cite{Chak}. 

In Fig. 1, we show the neutrino emissivity as a function of baryon density
at $B_m=0$ (see Eq. (9)) for the direct URCA process in nuclear matter
(denoted by NM) at an interior temperature $T=10^9$ K. Due to momentum 
conservation, the threshold density at which this process occurs, is at 
$n_t = 0.346$ fm$^{-3}$. The variation of emissivity with $n_b$ in presence
of magnetic field $B_m$ (see Eq. (8)) as seen in the figure may be explained
as follows: At very low densities $n_b\sim 0.35-0.73$ the field $B_m$ 
(as given in Eq. (12)) is rather small $\stackrel{<}{\sim} 10^{18}$ G, 
and consequently
a large number of Landau levels are populated. This gives essentially 
field-free results. At densities $n_b\geq 0.75$ fm$^{-3}$, the field is strong
enough to populate only the ground levels of both electrons and protons
\cite{Chak}, and would have pronounced quantization effects; the critical 
field for electron is $B_m^{(e)(c)} = 4.414\times 10^{13}$ G. 
The emissivity then rapidly increases with density and could have values as 
high as $\sim 2$ orders of magnitude larger than $B_m=0$ case at 
$n_b\approx 1.2$ fm$^{-3}$. Hereafter, $B_m$ saturates to a maximum of 
$5\times 10^{18}$ G so that for $n_b>1.2$ fm$^{-3}$, 
higher level states start to populate, and, as in the low density 
situation, results in field-free emissivity values. The central densities 
$n_c$ of neutron stars with maximum masses are also shown in Fig. 1 with 
(open circles) and without (solid circles) the magnetic field. 
For $B_m\neq 0$ star, $n_c = 1.448$ fm$^{-3}$ and thus falls above the 
kernel of enhanced emissivity leading to faster cooling compared to the 
field-free case. 

The neutrino energy losses from direct URCA processes of quark matter
composed of free $u$, $d$ and $s$ quarks and $e$ are estimated in the 
bag model. The current masses of the quarks are taken as $m_u=5$ MeV, 
$m_d=10$ MeV and $m_s=150$ MeV, and the bag constant as $B=250$ MeV fm$^{-3}$. 
In Fig. 1, we display the neutrino emissivity from the $\beta$-decay of 
$d$ and $s$ quarks at $B_m=0$ (see Eq. (10)) in quark matter (denoted by QM).
The $d$ quark $\beta$-decay reactions are kinematically allowed if 
$n_b\stackrel{>}{\sim} n_0$. At densities $n_b \simeq 0.85$ fm$^{-3}$ and
above when s quark decay is allowed, the emissivity is increased to about an
order of magnitude. This is 
caused by the large $s$ quark mass which allows
the momenta of the free particles to deviate appreciably from collinearity
which tends to increase the matrix element. It was, however, shown \cite{Iwa}
that by the inclusion of quark-quark interaction, the neutrino emissivities
from $d$ and $s$ quark $\beta$-decay are comparable in magnitude. The 
emissivities for the quark direct URCA processes in presence of the 
magnetic field (see Eq. (11)), remain virtually unaltered
from the field-free case due to the population of a large number of levels
in all the quark species. In either case, it is found that 
$\varepsilon_{\rm URCA}^{\rm QM} /\varepsilon_{\rm URCA}^{\rm NM} 
\stackrel{<}{\sim} 10^{-3}$.
{\centerline{
\epsfxsize=11cm
\epsfysize=13cm
\epsffile{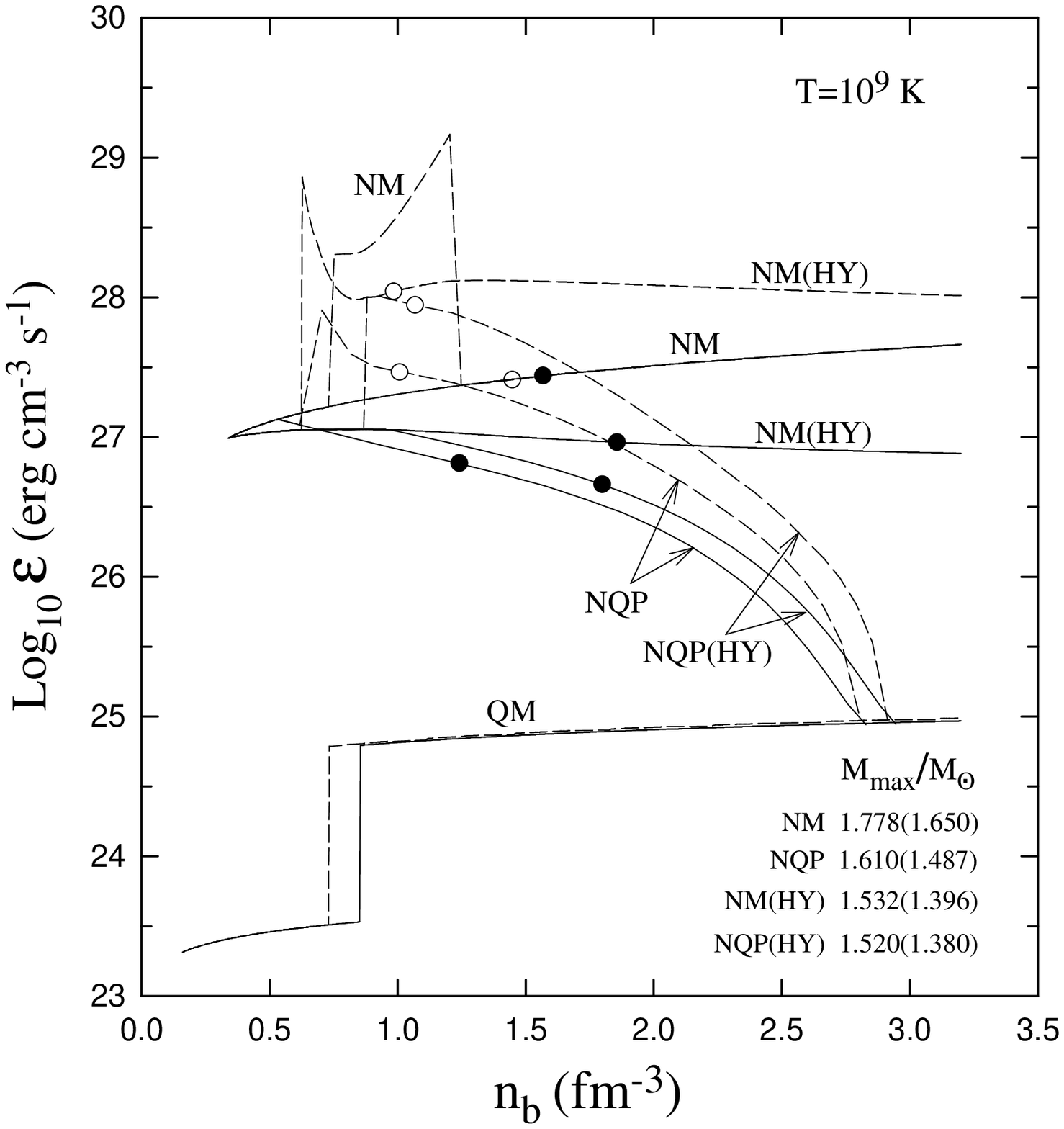} 
}}

\vspace{-2.2cm}

\noindent{\small{ FIG. 1. The neutrino emissivities as a function of 
baryon density from 
the direct URCA process for a magnetic field $B_m=0$ (solid line) and
for $B_m = 10^8-5\times 10^{18}$ G (dashed line) for: nucleons in 
nuclear matter (NM); quarks in quark matter (QM); a nucleon-quark phase 
transition (NQP); nucleons in nuclear matter with hyperons (NM(HY)); a 
nuclear matter with hyperons to quark phase transition (NQP(HY)). The 
maximum masses of the stars $M_{\rm max}$ with these various compositions 
are given for $B_m=0$ and those in the parentheses are for 
$B_m=10^8-5\times 10^{18}$ G. The corresponding central densities are 
indicated by solid and open circles, respectively.}}
\vspace{1.0cm}

In a realistic situation, if quarks at all exist, a star with increasing 
density from the surface to the center would have a pure nucleon phase at the 
inner crust and core with a possible pure quark phase at the center and a 
mixed nucleon-quark phase (NQP) in between. The mixed phase of nucleons and 
quarks is described following Glendenning \cite{Gle}.
The conditions of global charge neutrality and baryon number conservation 
are imposed through the relations $\chi Q^n + (1-\chi)Q^q = 0$ and 
$n_b = \chi n_b^n + (1-\chi)n_b^q$, where $\chi$ represents the fractional
volume occupied by the hadron phase. Furthermore, the mixed phase 
satisfies the Gibbs' phase rules: $\mu_p = 2\mu_u + \mu_d$ and $P^n = P^q$.
The neutrino energy loss rate in this phase is given by 
$\varepsilon_{\rm URCA}^{\rm NQP} = \chi\varepsilon_{\rm URCA}^{\rm NM}
+ (1-\chi)\varepsilon_{\rm URCA}^{\rm QM}$. The neutrino 
emissivities for the nucleon-quark phase transition are shown in Fig. 1
(denoted by NQP). For $B_m=0$ case, with the appearance of the quarks
at $n_b =0.533$ fm$^{-3}$, the emissivity decreases from the corresponding 
NM case. Apart from the reduced emissivity of the quark phase (which being,
however small at large $\chi$), the reduction in the chemical potentials
of the nucleons and electrons resulting from the requirement of the global 
charge neutrality and baryon number conservation conditions in the mixed 
phase, primarily causes the decrease in emissivity. For stars with 
$B_m\neq 0$, the emissivities in the mixed phase are enhanced, 
particularly in the regime dominated by nucleons (i.e. for $\chi>0.5$). 
The central densities of maximum mass stars fall within the mixed phase, and 
consequently such stars would have faster cooling than pure quark stars. 
The maximum mass NQP stars with and without magnetic 
field, however, have much smaller emissivities than that of the corresponding 
NM stars, while NQP stars with $B_m$ have nearly identical cooling as that of 
field-free NM stars even though their maximum masses are very distinct.

Since quark matter furnishes both baryon number and negative charge, 
intuitively, the trends exhibited by the emissivities for NQP stars may be 
anticipated by invoking strange baryons, namely  hyperons ($\Lambda$'s, 
$\Sigma$'s and $\Xi$'s). The $\beta$-equilibrium conditions then generalize 
to $\mu_i=b_i\mu_n-q_i\mu_e$, where $b_i$ and $q_i$ are the baryon number
and charge for the $i$th particle. Since the hyperons are more massive than 
the protons, the effect of the magnetic field on their direct URCA processes
is negligible. Because of the large uncertainties in the hyperon-nucleon 
interactions even at  nuclear density, for a conservative estimate of the 
emissivities we set the nucleon-meson and hyperon-meson coupling constants 
equal. Furthermore, the critical density
for nucleon direct URCA process is nearly identical to the hyperon 
threshold density in the relativistic mean field model, and the 
emissivities from the hyperon direct URCA processes are about 5-100 times 
less than that from the nucleons \cite{Pra}. Therefore, we shall present 
emissivity vs $n_b$ results only for the nucleon direct URCA process 
in presence of hyperons. This is shown in Fig. 1 and denoted by NM(HY). 
With the appearance of hyperons, the reduction in the chemical potentials 
of the nucleons and electrons required by the baryon number conservation 
and charge neutrality condition causes a substantial reduction of the 
emissivity compared to that from NM. In fact, with increasing
density when hyperon abundances grow rapidly, the emissivities gradually
decrease. For $B_m\neq 0$, only the ground Landau levels for
$e$ and $p$ are populated over a considerable density range in this matter.
Consequently, the emissivities with $B_m$ in NM(HY) stars are significantly
larger than that for the corresponding field-free stars.

Allowing now baryon  to quark phase transition, the emissivity displayed 
in Fig. 1 (denoted by NQP(HY)) for $B_m=0$ is larger than that from the NQP 
matter. This is caused by the delayed appearance of quarks in hyperon rich 
matter, so that the the total emissivity is primarily dominated by the nucleons. 
In presence of the field, the total emissivity of NQP(HY) matter is about 
an order of magnitude larger for the maximum mass star and therefore leads 
to faster cooling compared to the corresponding field-free star.

Within the non-relativistic (but interacting) approximation for the specific
heat $c_v$ and neutrino emissivity, the time for the center of a NM star to 
cool by the direct URCA process to a temperature $T_9$ at $B_m=0$ may be 
estimated to be $\Delta t = -\int(c_v/\varepsilon_{\rm URCA}^{\rm NM}) dT
\sim 10 T_9^{-4}$ s. In contrast, for $B_m\neq 0$, the NM star's center
cools faster with $\Delta t \sim 0.5T_9^{-4}$ s. By invoking quarks 
and/or hyperons, the decrease in emissivity is much more compared to that 
of the specific heat resulting in slow cooling of the stars center.
The typical time scale associated with the propagation of thermal signals 
through the outer core and the crust to the surface before the sudden 
temperature drop is quite high $\sim 1$ to 100 yr, depending on the 
crustal composition and relative sizes of the crust and the core and thus 
upon the equation of state. Therefore, it seems to be quite difficult to 
distinguish observationally from the effects of direct URCA process, the 
interior constitution of a star.

Throughout our discussion we have assumed that the electron is the only
lepton. If the triangle inequality $p_{F_p}+p_{F_\mu}\geq p_{F_n}$ is 
satisfied then nucleon direct URCA process with muons will occur; the 
threshold density for this process is higher than electrons since 
$m_\mu>m_e$. The $\beta$-equilibrium condition $\mu_e=\mu_\mu$ moreover
implies that the emissivity for URCA process with muons is same as that
for the corresponding process with electrons. In the present model the 
nucleon direct URCA processes are not permitted at densities $n_b<0.34$
fm$^{-3}$. In this outer core region, the dominant neutrino emission process 
are then the modified URCA processes of reaction (1) for which the 
emissivity is smaller by a factor $\sim (T/T_F)^2$ than the direct URCA
processes.

At certain densities and temperature $T<T_c\approx 10^8-10^{10}$ K, the
nucleon superfluidity may set in. The specific heat and direct URCA rate
are then reduced by a factor $\sim \exp(-\Delta/T)$, where $\Delta$ is the
larger of the neutron and proton gaps. The modified URCA rates are, however, 
reduced by a factor $\sim\exp(-2\Delta/T)$. In presence of a magnetic field, 
the superfluid protons are believed to form a
type II superconductor in the outer core within the density range 
$0.7n_0<n_b<2n_0$, and the estimated lower and upper critical magnetic 
fields are respectively $H_{c1} \sim 10^{15}$ G and 
$H_{c2} \sim 3\times 10^{16}$ G \cite{Bay}. With the choice of variation of
$B_m$ with $n_b$ (see Eq. (12)), the field is $10^{14}-10^{16}$ G at the 
bulk of the outer core and could form a superconducting region at $T<T_c$,
while the inner core and center with $B_m\sim 10^{18}$ G is in the 
normal state without superconductivity. 

In conclusion, the neutrino emissivities for all the cases studied here 
are found to be dramatically enhanced in a magnetic field compared 
to that from the non-magnetized stars. However, for certain stars 
unambiguous determination of the interior constituents may be difficult. 
There can be stars with the same composition, as for example the NM stars in 
a magnetic field but with slightly different masses of $1.55M_\odot$ and 
$1.60M_\odot$ having their central densities at 0.72 fm$^{-3}$ and 0.88 
fm$^{-3}$ residing below and within the kernel of fast cooling respectively, 
and thus have completely different emissivities. On the other hand, a NM star 
with $B_m=0$ and a NQP star in a magnetic field, though possessing different 
interior compositions, have nearly identical emissivities. 
It is also found that when pure nuclear matter is injected with nonleptonic 
negative charges, namely hyperons and quarks, the emissivities turn out to 
be smaller than that from the nuclear matter. It has been already demonstrated 
\cite{PraD} that nonleptonic negative charges cause a softening of the equation 
of state. We thus arrive at a general result that when matter contains 
nonleptonic negative charges, the maximum masses of the stars are smaller 
with a suppression of the neutrino emissivity than that of the pure nuclear 
matter with and without a magnetic field.

\end{document}